# The polarization of light and the Malus' law using smartphones

**Martín Monteiro[a], Cecilia Stari[b], Cecilia Cabeza[c], Arturo C. Marti[d],**

[a] Universidad ORT Uruguay; monteiro@ort.edu.uy

[b] Universidad de la República, Uruguay, cstari@fing.edu.uy

[c] Universidad de la República, Uruguay, cecilia@fisica.edu.uy

[c] Universidad de la República, Uruguay, marti@fisica.edu.uy

Originally an empirical law, nowadays Malus' law is seen as a key experiment to demonstrate the transverse nature of electromagnetic waves, as well as the intrinsic connection between optics and electromagnetism. In this work, a simple and inexpensive setup is proposed to quantitatively verify the nature of polarized light. A flat computer screen serves as a source of linear polarized light and a smartphone (possessing ambient light and orientation sensors) is used, thanks to its built-in sensors, to experiment with polarized light and verify the Malus' law.

**Smartphone-based experiments and optics**

In the last years, a great deal of smartphone-based experiments have been proposed in physics. Remarkably, experiments focusing on light and optics, and specially those using the ambient light sensor[1-2], have received little attention compared to those focusing on mechanics, oscillations or magnetism. Two exceptions are worth mentioning. In Ref.[1], the authors proposed a simple verification of the inverse square law using the light sensor of a smartphone or a tablet. In a different approach[2], the ambient light sensor has been proposed to indirectly measure distances and to analyze coupled springs undergoing oscillatory motions. Here, we focus on an experiment using the ambient light sensor which involves the polarization of light[3-9] and, in particular, the Malus' law.

Smartphones also gives us the ability of measuring simultaneously with various sensors. This is also a great advantage since it allows a great variety of experiments to be performed, even outdoors, avoiding the dependence on fragile or unavailable instruments. In previous works, the simultaneous use of two sensors like the gyroscope and the accelerometer was proposed to relate angular velocity, energy, centripetal and tangential acceleration[10-12]. In another experiment, the pressure sensor and the GPS were used in synchrony to find the relationship between atmospheric pressure and altitude[13].

In this work, we propose an experiment in which we take advantage of the capabilities of a smartphone to verify the Malus' law. The intensity of polarized light from a computer monitor is measured by means of the ambient light sensor with a tiny polarizer attached to it while the angle between the polarization and the polarizer is measured by means of the orientation sensor. The simultaneous use of these two sensors allows us to simplify the experimental setup and complete a set of measures in just a few minutes. The experimental results of the light intensity as a function of the angle shows an excellent agreement with Malus' law.

**Polarized light**

Light, as any other electromagnetic wave, nearly always propagates as a transverse wave, with both electric and magnetic fields oscillating perpendicularly to the direction of propagation (see a standard general physics textbook). The direction of the electric field is called the *polarization* of the wave. In a linearly polarized plane wave the electric field remains in the same direction. This pure state of polarization is called *linear polarization*. Natural light, (e.g., light radiated by an incandescent object) as a random mixture of waves with different polarizations, is unpolarized light, or more precisely, random polarized light.

According to conservation of energy applied to electromagnetic fields (Poynting's theorem), the energy flow (intensity or illuminance, *I*) associated to an electromagnetic wave (light) is proportional to the square of the amplitude of the electric field. When light interacts with matter its behavior is modified, mainly its intensity and its velocity. Moreover, some materials are able to modify light differently in each spatial direction. This is the case for instance of linear polarizers that can convert unpolarized light into linear polarized light. An ideal polarizer fully attenuates light polarized in one direction, and fully transmits light with the orthogonal polarization.

Consider a beam of linear polarized light incident over a polarizer. Let $\theta$ be the angle between the axis of the polarizer and the polarization of the incident light. The electric field that passes through the polarizer is the component in the direction of the axis, $E = E_0 \cos\theta$. Therefore, the intensity of the light passing the polarizer is

$$I = I_0 \cos^2\theta \quad (1)$$

where $I_0$, is the intensity of the light before the polarizer. Equation 1 is the so called **Malus' law**, named after the French physicist Étienne-Louis Malus, who discovered optical polarization in 1808.

**The experiment**

In this experiment, a source of polarized light, a polarizer, a photometer and a way to measure angles are needed. The source of linear polarized light is a flat computer monitor (or LCD TV screen) in plain white color[4-6]. The ambient light sensor of LG-G3 smartphone, located near the front camera (Fig. 1, left panel) is used as a photometer and the orientation sensor is used to measured the angles. A small piece of polarizer (a square of about 1 cm side) from an old calculator's display was placed over the ambient light sensor as shown in the central panel of Fig. 1.

The ambient light sensor works as a linear photometer and measures the illuminance, *i.e.,* the total luminous flux incident on a surface, per unit area whose unit is lux. However, in the Malus' law, the relevant variable is the irradiance (or light intensity, *i.e.*, the total power received by a surface per unit area measured in $W/m^2$)[14]. The illuminance, contrarily to the irradiance, considers the fact that human eyes' are more sensitive to some wavelengths than others, and, consequently every wavelength is weighted differently. In our experimental setup, since the spectrum of the light source does not change, the irradiance is proportional to the illuminance.

The angle between the polarized light from the screen and the polarizer attached to the smartphone is obtained using the smartphone's orientation pseudo-sensor[15]. This sensor provides the three necessary angles to determine the orientation of the smartphone: pitch, roll and azimuth. In particular, the pitch is the angle between the horizontal direction and the y-axis of the smartphone.

The *app* Physics Toolbox[16] is used to register simultaneously the illuminance and the angle. This app has an option, *Multi report*, in which we can choose the sensors that we are going to use. In this case we checked the boxes of light and orientation sensors. With the app started, the smartphone with the polarizer is placed in front of the monitor in upright position as indicated in Fig. 1 (right panel). Note that, the distance does not change and the light intensity is constant.

One important aspect of the experimental setup is that the smartphone pitch-angle be coincident with the angle between the polarization and the polarizer as illustrated in the Fig. 2. To do so, the experiment starts by placing the smartphone upright over the screen, with a pitch angle of -90º. Next, keeping the smartphone upright, the polarizer is rotated looking for a minimum in the light intensity. In this position, the polarizer is fastened with a tiny piece of tape over the light sensor. As a result, the axis of the polarizer is perpendicular to the polarization of the light from the screen, as indicated in Fig. 2 (left panel).

Afterward, we start collecting data with the *app* and the smartphone is gently rotated in front of the screen completing at least a quarter of revolution. Note that, due to the symmetry of the Malus' law, collecting data corresponding to a more large range will produce a repetition of the data. Once the data is recorded, the Physics Toolbox app saves a *csv* file that can be downloaded to a PC or tablet and analyzed using appropriate software. *csv* files, composed of several columns (displaying the variables from the sensors chosen) separated by a comma character (or another character according to your local configuration) are very easy to deal with. In this experiment we use only the columns corresponding to the pitch angle and illuminance.

**Results and Conclusion**

The experimental values directly obtained from the smartphone are shown in Fig. 3, together with the theoretical prediction of the normalized illuminance as a function of the angle between the polarization and the polarizer. Experimental data was fitted according to a logarithmic linearization of Eq. 3, as shown in the inset of Fig. 3, where the slope of the linear fit indicates the exponent in the Malus' law.

All in all, the experimental values are in excellent agreement with the expected ones, and we conclude that thanks to the simultaneous use of two less exploited sensors of a smartphone it is possible to verify the Malus' law in a way that is very accessible to the students, promoting autonomy and engagement. Moreover, more experiments using smartphones and polarizers can be devised, for example with waveplates or retarders (like half-wave plates and quarter-wave plates) or circular polarizers such as those used in some 3D movie theaters or even with inexpensive cellophane tape to explore birefringence[17].

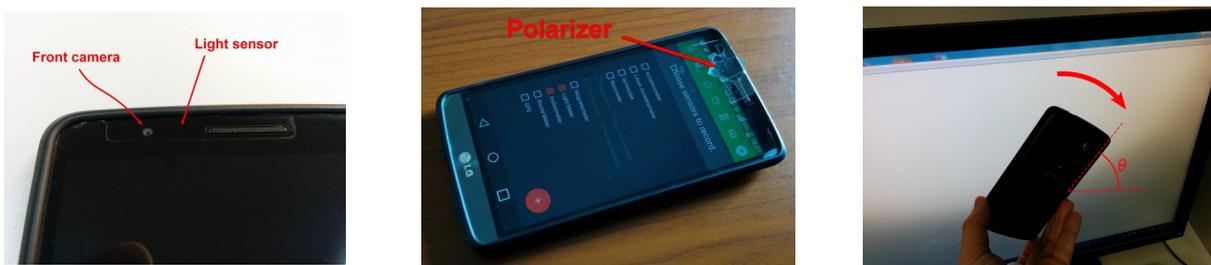

Figure 1. The ambient light sensor (left panel), a polarizer above the ambient light sensor (center panel) and a smartphone placed over a vertical screen (right panel). While the smartphone is gently rotated, the ambient light and the angle is simultaneously registered by the sensors.

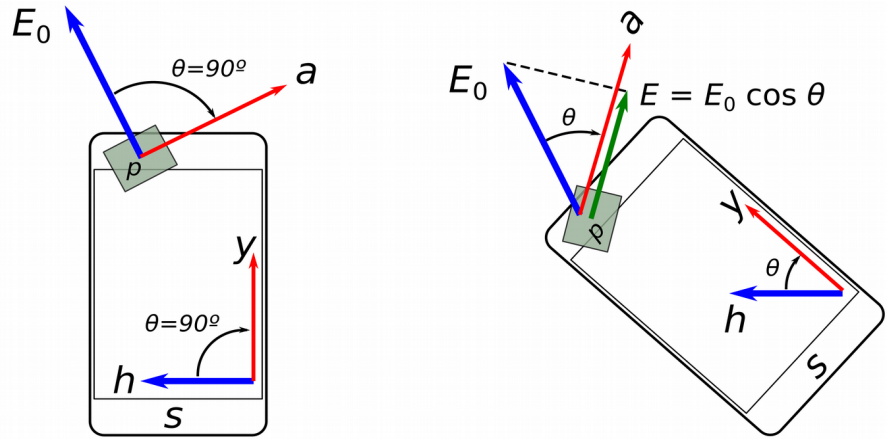

Figure 2. The smartphone, *S*, with the polarizer, *p*, attached over the ambient light sensor viewed from the computer screen. Initially (left panel), the smartphone's *y*-axis is aligned with the vertical direction (pitch angle = -$\theta$ = -90º) and the polarizer axis, **a**, is adjusted to be perpendicular to the electric field, $E_0$, of the polarized light from the computer screen. While the smartphone is being rotated (right panel), the resulting electric field that passes through the polarizer and is incident upon the light sensor is $E = E_0 \cos\theta$.

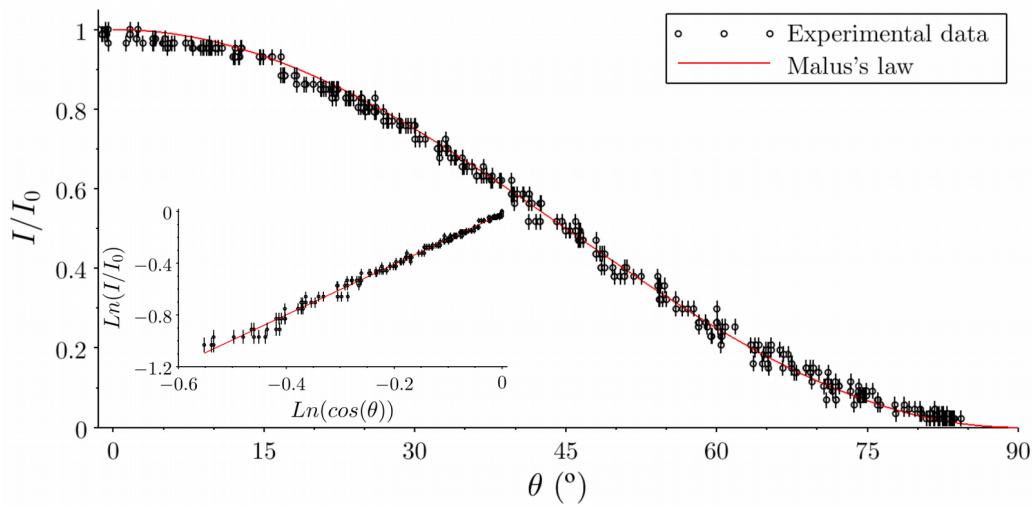

Figure 3. Experimental data (circles) of the normalized intensity of light as a function of the angle between the polarization and the polarizer, and the theoretical expression of the Malus's law (red line). The light intensity is normalized to the maximum of the transmitted light, $I_0$. The inset shows the analogous logarithmic plot as a function of the cosine of the angle and a linear fit (red line) whose slope coefficient, $1.96 \pm 0.02$, corresponds to the exponent in the Malus' law. The regression coefficient of the linear fit is $R^2 = 0.99316$.